\documentclass[10pt,twocolumn]{article}
\usepackage{times}
\usepackage{graphicx}
\usepackage{xcolor,colortbl,xspace}

\usepackage{ifthen}
\newboolean{publicversion}

\setboolean{publicversion}{false}

\ifthenelse{\boolean{publicversion}}{
  \newcommand{\grumbler}[3]{}
}{
  \newcommand{\grumbler}[3]{\xspace\textcolor{#3}{\bf #1: #2}}
}

\begin{document}

\title{Towards Software-Defined Data Protection: \\ GDPR Compliance at the Storage Layer is Within Reach \\ \normalsize [Vision Paper]}

\author{Zsolt Istv\'an\\IMDEA Software Institute, Madrid\\\emph{zsolt.istvan@imdea.org}\and Soujanya Ponnapalli\\University of Texas, Austin\\\emph{soujanya.ponnapalli@utexas.edu} \and Vijay Chidambaram\\University of Texas, Austin and VMWare\\\emph{vijay@cs.utexas.edu}}

\date{}
\maketitle

\begin{abstract}
  Enforcing data protection and privacy rules within large data processing
  applications is becoming increasingly important, especially in the
  light of GDPR and similar regulatory frameworks. Most modern data
  processing happens on top of a distributed storage layer, and
  securing this layer against accidental or malicious misuse is
  crucial to ensuring global privacy
  guarantees. However, the performance overhead and the additional complexity for this is often assumed to be significant -- in this work we describe a path forward that tackles both challenges. We propose ``Software-Defined Data Protection'' (SDP),
  an adoption of the ``Software-Defined Storage'' approach to
  non-performance aspects: a trusted controller translates company and
  application-specific policies to a set of rules deployed on
  the storage nodes. These, in turn, apply the rules at line-rate but
  do not take any decisions on their own. Such an approach decouples
  often changing policies from request-level enforcement and allows
  storage nodes to implement the latter more efficiently.

  Even though in-storage processing brings challenges, mainly because
  it can jeopardize line-rate processing, we argue that today's Smart Storage 
  solutions can already implement the required functionality, thanks to the
  separation of concerns introduced by SDP.  We highlight the challenges that 
  remain, especially that of trusting
  the storage nodes. These need to be tackled before we can reach widespread
  adoption in cloud environments.

\end{abstract}

\section{Introduction}

Our online presence has been generating unprecedented
amounts of data, a large portion of which are personally identifiable and
hence prone to misuse. Even though companies have long used different
access control and encryption techniques to secure the information
they collect, with the emergence of
regulatory frameworks such as GDPR in the EU~\cite{gdpr-regulation} and CCPA in
California~\cite{ccpa},
there is a need to homogenize and perhaps standardize these techniques to
enforce rules at all levels of application
stacks~\cite{shastri2019seven,schwarzkopf2019position}. In this work,
we focus on enforcing privacy rules at the storage level, but tracking
data through processing steps of large-scale applications
or when shared with third parties, is just as
important. Nonetheless, analysis of GDPR finds that more
than 30\% of the articles enforcing data protection are related to 
storage~\cite{shah2019analyzing}. Maintaining strict compliance is challenging 
because it requires computational resources beyond the capacity of 
storage nodes and could slow down data access.

There have been two emerging trends that bring an optimistic outlook 
to enforcing privacy rules directly in the storage layer. First, disaggregated
architectures in the cloud and datacenters are increasingly offering 
some form of in-storage processing~\cite{samsungsmartssd,do2019programmable} and flexible data
management~\cite{jun2016bluedbm,istvan2017caribou} with high-speed
network connectivity. Second, as an effort to keep the management,
configuration and monitoring of a large number of storage nodes
scalable, Software-Defined Storage (SDS) has been
proposed~\cite{thereska2013ioflow} -- albeit the goal of existing SDS systems
is almost exclusively to guarantee performance isolation and service
levels in distributed multi-tenant settings~\cite{macedosurvey}.

In this vision paper, we propose \emph{Software-Defined Data Protection (SDP)}
that decouples policy interpretation and decision making (control plane) from
request processing (data plane), reducing the complexity of the
functionality required inside the storage nodes that often have
limited hardware resources. As a result, SDP makes it feasible
to implement complex policies, such as GDPR, with state-of-the-art smart
storage devices that incorporate several low-power CPU cores and programmable
fabric (FPGAs)~\cite{samsungsmartssd}. Furthermore, the logically centralized control plane ensures 
that regardless of the physical location of the storage nodes, the same 
privacy rules apply, keeping behavior consistent even across datacenters.

Somewhat surprisingly, when adopting the SDP approach, GDPR-compliant storage, to a large extent,
can be already provided with existing ``building blocks''.
We sketch how these could assemble into a high bandwidth
pipeline within the storage node. We also highlight the most important remaining 
challenge in the cloud context, namely adding Trusted Execution
Environments (TEEs)~\cite{jauernig2020trusted} inside the storage nodes 
that would allow for remote attestation of the firmware by the controller.

\smallskip

\smallskip
\noindent
\hspace{-2mm}
\colorbox{blue!10}{
\begin{minipage}{0.48\textwidth}

To summarize, \textbf{our vision is that using in-storage computation together with a control-plane/data-plane separation will allow enforcing GDPR at line-rate}, without requiring storage nodes to be implemented with large, power-hungry, servers. The proposed Software-Defined Data Protection (SDP) can achieve this goal by re-purposing hardware building blocks that have been extensively studied in different contexts, and through this, bring GDPR-compliance to storage at little added cost. By sketching how a working solution would look like, \textbf{we identify the one missing piece in the state-of-the-art needed to make our vision reality: the possibility of using Trusted Execution Environments inside storage nodes} with heterogeneous hardware and near-data processing.

\end{minipage}
}


\section{Background and Related Work}

\subsection{Implications of GDPR in the Cloud and on Storage}

The General Data Protection Regulation (GDPR) outlines the rights and responsibilities
of the companies handling the personal data of EU citizens. GDPR, in 99 articles and
173 recitals, regulates the entire life cycle of personal data, from its collection to
its deletion.


GDPR outlines that it is the responsibility of a company to use third-party services
that do not violate GDPR standards. A vast majority of companies today rely on cloud
services providers, for their infrastructural needs, making GDPR compliance in cloud
environments a necessity. Therefore, rich related work proposes frameworks enabling 
cloud users to develop GDPR-compliant
applications~\cite{Rios2019ServiceLA}, along with studies that outline the challenges
faced by cloud users in the face of GDPR~\cite{inproceedings}, and the support for the
right to be forgotten in hybrid clouds~\cite{kelly2020achieve}.


The impact of GDPR on storage systems~\cite{shah2019analyzing} has been broadly studied
previously. GDPR's impact on databases~\cite{10.14778/3384345.3384354} and how
databases, by design, can comply with GDPR~\cite{schwarzkopf2019position,kraska2019schengendb}
are also analyzed. Recent work investigates how systems can violate
GDPR~\cite{shastri2019seven}, proposes benchmarks~\cite{10.14778/3384345.3384354}
and provides tools~\cite{8933680} that test GDPR-compliance, and that explore the benefits of
Trusted hardware to prove GDPR-compliance~\cite{Mazmudar2019MitigatorPP}.



The six features required from a GDPR-compliant storage system~\cite{shah2019analyzing} are as follows:

\noindent
\textbf{1. Deletion}. GDPR introduces the right to be forgotten, allowing users to
demand the deletion of personal data. GDPR also states that personal data cannot be
stored beyond its purpose in its storage limitation clause, requiring storage
systems to support user data deletion.

\noindent
\textbf{2. Logging and Monitoring}. With GDPR, companies are vested with the
responsibility of detecting potential data breaches, informing users about those data
breaches, and proving their compliance. GDPR also provides users with the right to
access, allowing them to request with whom and why their personal data is shared.
These rights and requirements necessitate some form of logging and monitoring at the
storage layer.

\noindent
\textbf{3. Metadata and Secondary Indexes}.As GDPR requires personal data to be associated with a
specific purpose, storage has to accomodate for additional metadata. As a performance optimization, 
secondary indexes that categorize data as per purposes can be useful. 

\noindent
\textbf{4. Fine-grained Permissions}. With GDPR's right to object, users can object
to using their personal data for specific purposes. Further, with the purpose
limitation clause, GDPR disallows companies to process user's data beyond its
purpose and without appropriate security measures. To comply with these clauses,
fine-grained permission checks and access control becomes crucial at the storage layer.

\noindent
\textbf{5. Encryption}. GDPR mandates that personal data should be protected against
accidental loss or damage and should be incomprehensible to any person unauthorized to
access it. Encrypting personal data is a suitable measure to comply with this clause.

\noindent
\textbf{6. Location Control}. GDPR requires companies to adhere to its standards,
independent of the geographical location of where personal data is stored.

\noindent
\textbf{7. Data Integrity}. We further identify that GDPR requires storage systems
to resist or detect malicious activities that compromise the integrity and
confidentiality of personal data. We propose using checksums or Merkle tree-based data
integrity checks to comply with the integrity and confidentiality clause of GDPR.

\subsection{Emerging Smart Storage Devices.}

Two kinds of smart storage devices are developed recently: those which incorporate small CPU cores, such as ARM cores~\cite{jo2016yoursql,koo2017summarizer}, and those which deploy specialized compute elements or FPGAs~\cite{jun2016bluedbm,istvan2017caribou,samsungsmartssd,bauer2018programmable}. The former category offers more flexibility, but as explored in depth in the work of Koo et al.~\cite{koo2017summarizer}, it can be difficult to predict the performance of the in-storage computation, and this can lead to performance degradation. Those solutions that rely on FPGAs and similar specialized hardware are a better match for network-focused implementations because they can guarantee network line-rate processing by design. To sample from our previous work, we have shown, among other operations, that it is possible to implement line-rate hash-tables~\cite{istvan2017caribou}, deduplication schemes~\cite{kuhring2019specialize}, in-storage filtering with regular expressions~\cite{istvan2016runtime} on FPGA-based distributed storage. These operations are similar in nature to those required for building an SDP system (discussed in Section~\ref{sec:required-func}).


\begin{figure}[t]
\centering
\includegraphics[width=\linewidth]{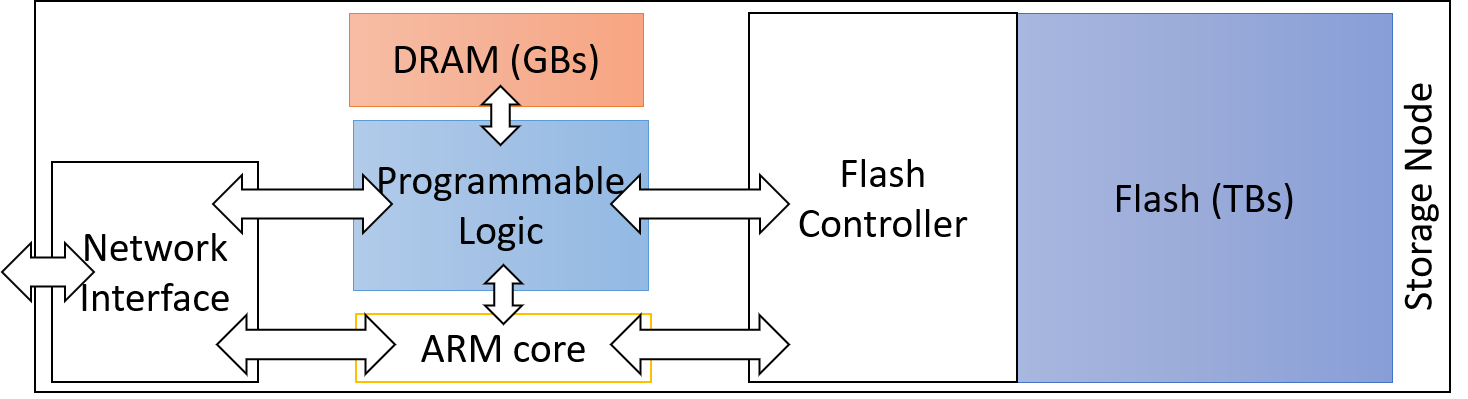}
\vspace{-1em}
\caption{\label{fig:inside-node} Emerging Smart Storage solutions incorporate both ARM cores and Programmable Logic (FPGAs) offering both flexibility and predictable performance.}
\end{figure}

Overall, the SDP approach is beneficial in both cases because it reduces the complexity of the code running inside the storage nodes and allows developers to think of the enforcement as pipeline stages.
We sketch our proposal targeting hardware platforms with a combination
of ARM cored and FPGA fabric (as depicted in
Figure~\ref{fig:inside-node}), such as the Fidus
Sidewinder-100~\cite{bauer2018programmable} or the Samsung
SmartSSD~\cite{samsungsmartssd,chapmancomputational}. The presence of
both a general-purpose, albeit low power, CPU core, and a
re-programmable specialized hardware element ensures that the devices
can be managed easily in a cloud setting and, at the same time, can
deliver high-performance behavior for privacy-related processing.

\begin{table*}[t]
\small
\centering
\begin{tabular}{|l|l|l|l|}
\hline
\rowcolor{lightgray!30}
\textbf{No.} & \textbf{GDPR article} & \textbf{Required functionality} & \textbf{Impacts mostly} \\
\hline
\rowcolor{yellow!20}
5.1 & Purpose limitation (data collected for specific purpose) & Fine-grained permissions  & Storage, Controller \\
\rowcolor{yellow!20}
21 & Right to object (data not used for objected reason) & Fine-grained permissions & Storage, Controller \\

\rowcolor{orange!20}
5.1 & Storage limitation (data not stored beyond purpose) & Deletion & Controller \\
\rowcolor{orange!20}
17 & Right to be forgotten & Deletion & Controller \\

\rowcolor{blue!20}
15 & Right of access by users & Metadata (and Secondary indexes) & Storage \\
\rowcolor{blue!20}
20 & Right to portability (transfer data on request) & Metadata (and Secondary indexes) & Storage \\

\rowcolor{green!20}
5.2 & Accountability (ability to demonstrate compliance) & Logging and Monitoring & Storage, Controller \\
\rowcolor{green!20}
30 & Records of processing activity & Logging & Storage \\
\rowcolor{green!20}
33, 34 & Notify data breaches & Logging and Monitoring & Storage, Controller \\

\rowcolor{red!20}
25 & Protection by design and by default & Encryption & Storage \\
\rowcolor{red!20}
32 & Security of data & Encryption and Access control & Storage \\

\rowcolor{gray!10}
13 & Obtain user consent on data management & High level policy$^\dagger$ & Controller \\
\rowcolor{gray!10}
46 & Transfers subject to safeguards & Location control$^\dagger$ & Controller \\
\hline
\end{tabular}
\caption{\label{tbl:gdpr-req}This table summarizes the GDPR articles relevant to storage and the high level functionality that the storage nodes and SDP controller need to fulfill those. Functionality outside of the scope of this paper is marked with~$^\dagger$.}
\end{table*}

\section{Software-Defined Data Protection}

SDP targets cloud and datacenter use-cases where data is being stored and processed within the context of a large corporation with several applications but governed by a single set of privacy rules. In line with DGPR requirements, we assume that user data can be identified explicitly through a universal \emph{user identifier} and a \emph{purpose identifier}. These identifiers are valid across applications of the company, and the SDP controller has a global view of which application can access what \emph{purpose}s. All \emph{user}s within a \emph{purpose} are accessible to the application, unless consent has been revoked.

\begin{figure}[t]
\centering
\includegraphics[width=\linewidth]{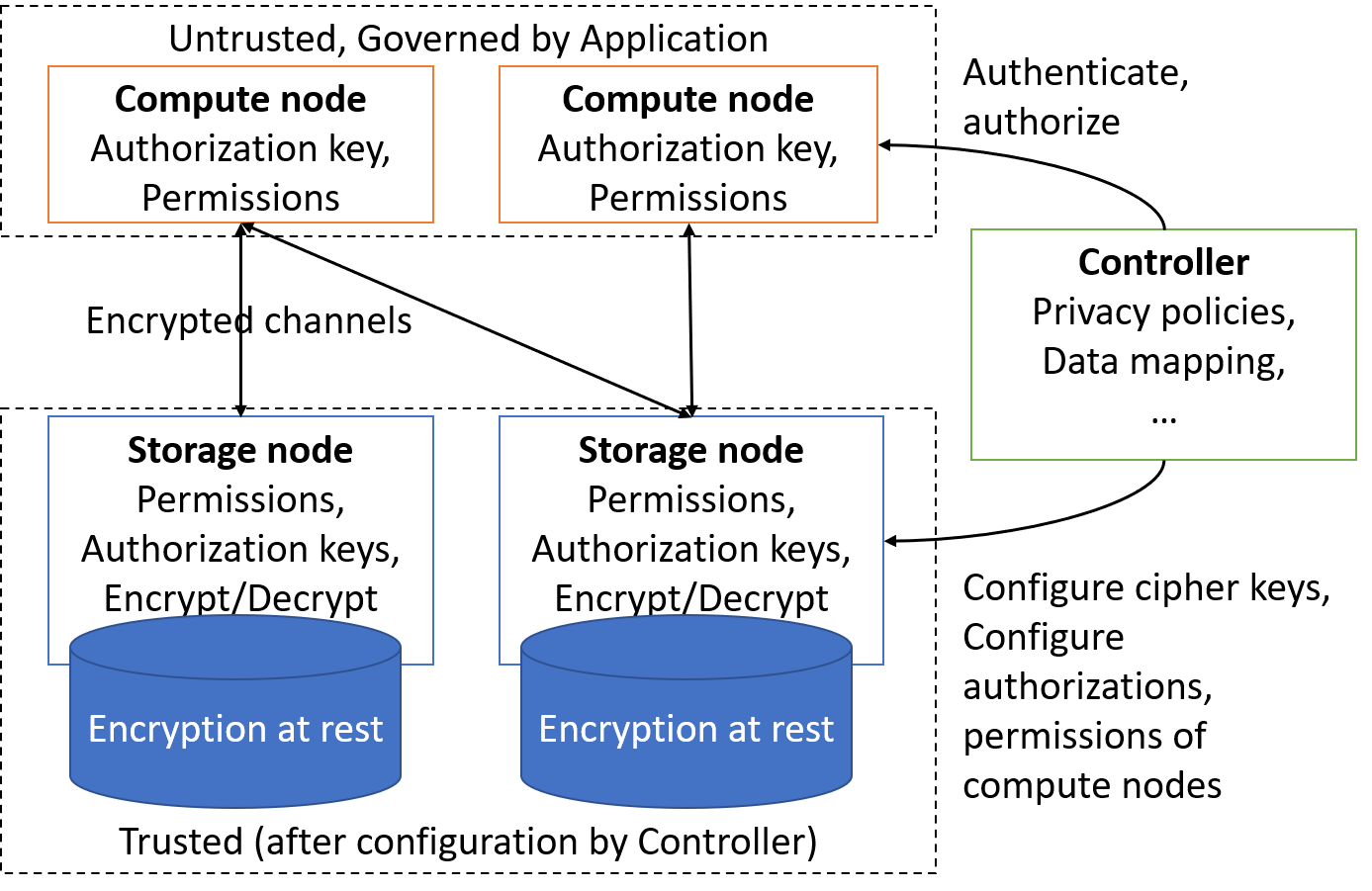}
\vspace{-1em}
\caption{\label{fig:sdp-overview} We propose Software-Defined Data Protection as a design approach to decouple policies from storage node implementation.}
\end{figure}

Figure~\ref{fig:sdp-overview} shows the three types of nodes in SDP: \emph{storage}, \emph{processing}, and \emph{controller}. \textbf{Storage nodes} implement a general-purpose key-value store (KVS) interface on binary data, can be shared across applications, and all communication with them happens over an encrypted channel (e.g., TLS). 

Applications run on one or several \textbf{processing nodes} and are managed under the governance of the developers. Applications have to register with the controller before accessing data but once granted access, can carry out most of their operations with the storage nodes.

The \textbf{controller node} is logically centralized and is trusted by both the storage and processing layers and is controlled by the company's data officers. It interprets policies and maintains data mappings and is used to bootstrap storage nodes, authenticate application nodes, and manage their permissions. Storage nodes are trusted once configured by the controller (see Section~\ref{sec:future-challenges} for how this could be ensured in practice in the cloud). As we explain in the following subsections, once authenticated and configured, in common case neither the application nor the storage nodes have to communicate with the controller. For this reason, we do not consider the controller as an obvious bottleneck for performance. 

\subsection{Required Functionality}
\label{sec:required-func}

In Table~\ref{tbl:gdpr-req}, we summarize the GDPR articles relevant to storage and what type of functionality is required for their fulfillment. We also indicate for each aspect, whether the SDP storage nodes (data plane) or the controller (control plane) would bear most of the required complexity. Naturally, even if the storage node performs almost all the computation, for instance, as is the case with Encryption, the controller will still have to bootstrap the nodes.

\begin{figure}[t]
\centering
\includegraphics[width=\linewidth]{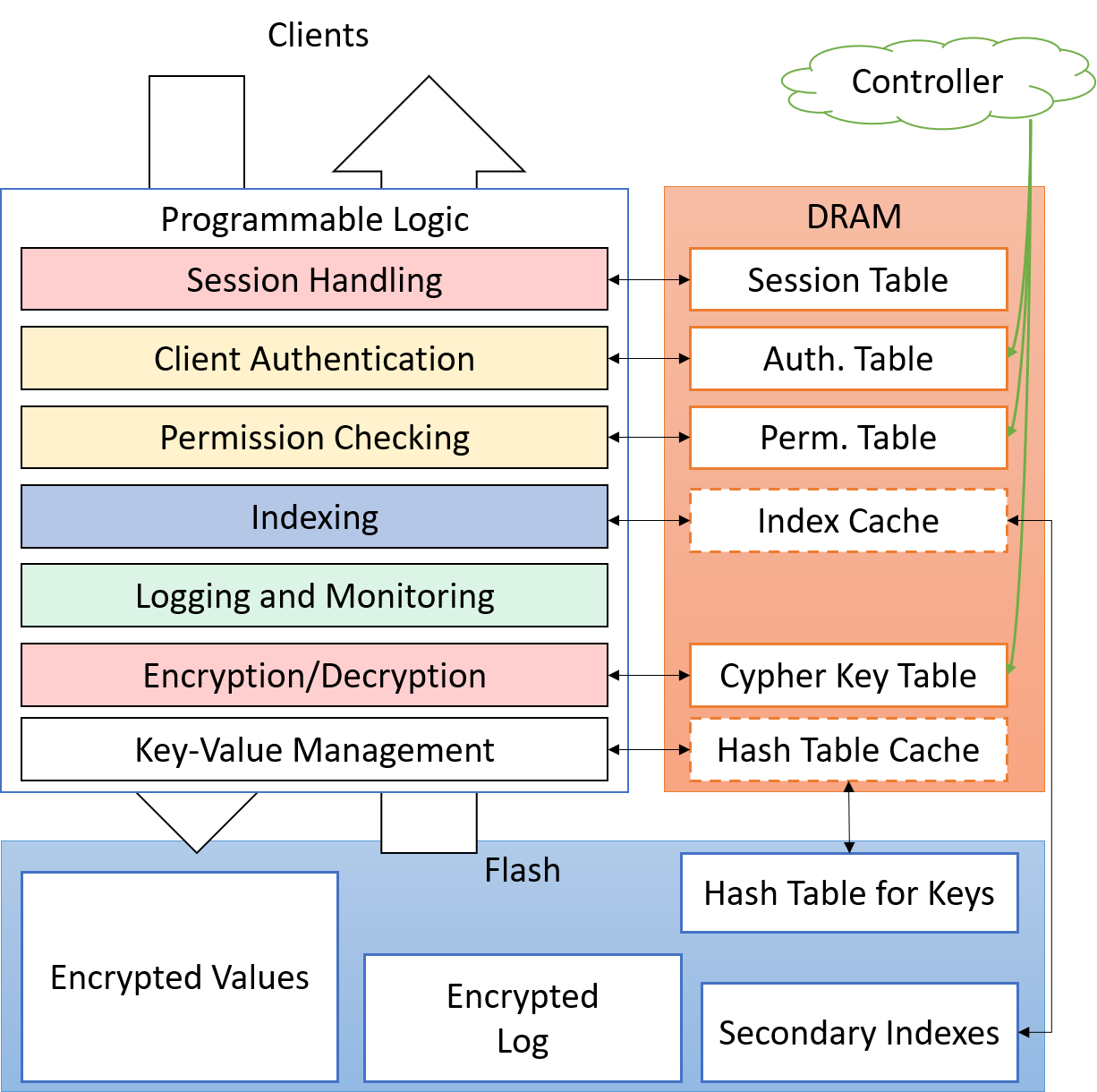}
\vspace{-1em}
\caption{\label{fig:pipeline} SDP enables separation of concerns, and as a result, policy enforcement can be laid out as pipeline within the storage nodes. The SDP controller configures and manages the nodes from the outside.}
\end{figure}

In the following, we discuss in more detail how the required functionalities can be implemented with state of the art modules in hardware (FPGA) as part of an SDS-inspired pipeline on the storage nodes. It is possible to implement most of the functionalities independent of each other (Figure~\ref{fig:pipeline}), and in future systems, non-performance-critical steps could be moved to low power CPU cores (for instance, Authentication that is only carried out once per session), achieving this way heterogeneous processing.

\smallskip \noindent \textbf{Encryption.}
In the following, we sketch how encryption can be implemented inside the storage layer in a way that \emph{a)}~relies on existing schemes and best practices and \emph{b)}~makes it possible to map the key-management operations to the SDP scheme we propose. 

Data needs to be encrypted both at rest and on the move.
We envision a system where clients (processing nodes) receive plain-text tuples over encrypted channels. Block ciphers underlying TLS have been shown to work well on FPGAs~\cite{caulfield2016cloud,hodjat200421} reaching throughputs high enough to saturate even 40Gbps links. 

Persistent data on flash is always encrypted, assuming industry-standard,
symmetric-key cryptography (e.g., AES).
The storage nodes must not persist cipher keys. Instead, they have to
remain in memory and need to be configured and managed by the SDP
controller at run-time. This forbids unauthorized access to the drives
and prevents leaks. One difference to traditional encrypted storage is
that in the GDPR context it can be beneficial if each tuple is encrypted 
with a key specific to the \emph{user} whose data it represents (or even the
\emph{user-purpose} combination) because, as later explained in the \emph{Deletion} subsection, this enables 
quick logical deletion of data.


A side-effect of having multiple (de)encryption keys is that each
client request will have to retrieve a different one. Since
requests encode \emph{user} and \emph{purpose} explicitly, the storage
node can use this to lookup the cipher key in an internal ephemeral
cipher key table (KT). It is important that this table can sustain
high access rates because, in contrast to other meta-data structures
described later, this one has to be accessed on every incoming client
request.

One challenge of creating fast hash tables on specialized hardware is
that it is unlikely that cipher keys will fit on on-chip SRAM
memory and will have to spill into off-chip DRAM. There is recent work
in the context of high-performance key-value stores built on
FPGAs~\cite{istvan2017caribou,li2017kv} using DRAM, which can be
adequate for this purpose.

\smallskip \noindent \textbf{Fine-grained Permissions.}
Beyond the question of how clients can reach storage devices (solved
by SDS), an authentication step is necessary. Authentication matches a client's
identity to a set of permissions (read/write/insert rights per
\emph{purpose}). To carry out authentication, an Authentication Table
(AT) holds the public keys belonging to applications. A Permission
Table (PT) stores the mapping of identities to permissions. Both
tables are ephemeral and are populated and managed by the
controller. In most workloads, the number of \emph{purposes} (e.g.,
number of internal applications) will be orders of magnitude smaller
than that of individual \emph{users}. Therefore the PT can be
represented compactly, perhaps even on on-chip caches.

Permissions in the PT are orthogonal to the presence/absence of cipher
keys in the KT: even if a client has the right to read all key-value
pairs belonging to a \emph{purpose}, only those for which the storage 
device holds a cipher key in the KT can be successfully
read. The same holds for inserting tuple belonging to a new \emph{user} or
\emph{purpose}. The SDP controller has to first insert the
corresponding entries in the KT. It is important to note that permissions 
in themselves do not forbid applications
to, for instance, make copies of data under bogus \emph{user} keys. 
For this reason, end-to-end information tracking is required, as we 
highlight this in Section~\ref{sec:future-challenges}.

\smallskip \noindent \textbf{Metadata and Indexing.}
By GDPR regulations it has to be possible to
retrieve all tuples belonging to a specific \emph{user} or
\emph{purpose}. This can be achieved by relying on meta-data stored
with the key-value pairs (and by enforcing a naming scheme of tuples).

While not strictly necessary, if such reads will occur often, it can
be beneficial to maintain secondary indexes for performance
reasons. Even though write operations will become more expensive,
these data structures can be used to avoid scanning TBs of data to
find a specific \emph{user}'s entries. For this purpose, there are
already FPGA-based key-value stores with low cardinality secondary
index~\cite{istvan2017caribou} and by using the same hash table
approach as for permissions, etc., higher cardinality cases could be
also handled.

\smallskip \noindent \textbf{Logging and Monitoring.}
Logging is important for ensuring auditability of the storage layer and can be implemented at various granularities. The storage node will persist an encrypted log (the key for the log is configured at run-time by the SDP controller) and implement some form of integrity check at the tuple level. While we foresee no challenges with the former task, the latter might require further investigation. Increasing the efficiency of data structures that ensure the integrity storage are still a topic in exploration~\cite{raju2018mlsm,ponnapalli2019scalable,bailleu2019speicher}.

For monitoring, the storage node has to notify the SDP controller of any request that failed any of the validation steps outlined above or has retrieved a key-value pair whose decryption key is missing. These events allow the controller to take adequate action, rectifying mis-configuration, revoking permissions, etc.

\smallskip \noindent \textbf{Deletion.}
In this work we focus on logical deletion of \emph{user} data, since physically destroying all copies, and proving this to a third party, is orthogonal to our goals and a challenge in itself~\cite{mitra2006secure,kim2020evanesco}. If using an encryption scheme with a single key, the SDP controller can delete tuples belonging to a \emph{user} and \emph{purpose} by delegating this task to the storage device that will either scan the tuple space or use a secondary index. But in case an encryption scheme with multiple keys is used, deletion can be performed more efficiently by simply removing the corresponding cypher keys from the KT of the storage nodes. As a result, deleted tuples will not be accessible any more as plain-text, and without the cipher keys, cannot be decrypted. This approach is in-line with existing practices~\cite{boneh1996revocable,zhu2005fossilized}.

Depending on the encryption scheme chosen, the logical delete operation could be performed with a single request to the storage node or in linear time. If \emph{(1)} the encryption of tuples is only based on the \emph{user} they belong to, then the controller can remove \emph{all} tuples of a user in constant time but will have to rely on secondary indexes to physically remove all tuples belonging to a \emph{purpose}. As an alternative, \emph{(2)} distinct cypher keys can be generated for each \emph{user:purpose} pair. This allows more fine-grained deletions and, even though not constant time, deleting all data of a \emph{user} is linear in the number of \emph{purposes}. As a third option, \emph{(3)} it is also a possibility to derive cypher keys on the fly from two separately stored parts for \emph{user} and \emph{purpose}. This enables constant time deletions of either \emph{all} data of a \emph{user} or \emph{all} data belonging to a \emph{purpose}, but more fine-grained deletions require costly scans in the storage. Regardless of the choice of the scheme, the functionality on the storage nodes changes very little because most of the complexity is handled by the controller.

\subsection{Features and Questions out of Scope}

\textbf{Location:} The mapping of \emph{user} and \emph{purpose} to
the actual storage devices is carried out by the controller following
either general purpose sharding strategies or depending on the privacy
policies at the high level. GDPR, however, mandates that regardless of the
physical location of data belonging to a \emph{user}, the same rules apply
to it. The logically centralized nature of the SDP controller is essential
for ensuring rule consistency at scale.

\smallskip\noindent
\textbf{High level policies:} The question of how company- and application-wide policies are written, managed and translated to SDP rules is out of the context of this paper. There is rich related work~\cite{krahn2018pesos,upadhyaya2015automatic,wang2019riverbed,sen2014bootstrapping} which demonstrates how to translate high level policies to compliant queries in databases (or compliant accesses in data storage layers) and we believe that they could be layered on top of SDP since our proposal envisions functionality which is a superset of such proposals. 

\smallskip\noindent
\textbf{Fault Tolerance by Replication:} For simplicity, in our discussion we assumed that each piece of data resides inside a single storage node. In a real system, however, replication will be required to ensure fault tolerance. As highlighted by the above two points, the task of setting up and controlling replication is external to our proposal and can be tackled by numerous existing schemes. Nonetheless, there is work on performing transparent, line-rate, replication of the KVS running on the FPGAs~\cite{istvan2016consensus} that could be easily adapted to be managed by the Controller.

\smallskip\noindent
\textbf{Performance of the Controller:} In our SDP vision, the Controller is required to actively participate only in a subset of operations. After carrying out initial authentications and configuring encryption keys and permissions, it is seldom accessed by either the processing nodes (application) or the storage nodes. Once exception is when data belonging to a new user or purpose is inserted for the first time. Such operations, however, are less common than regular reads and updates of existing data. Nonetheless, it is likely that the Controller will have to be implemented as a logically centralized by physically decentralized solution, to be able to keep up with the workloads of large enterprises.

\section{Future Challenges}
\label{sec:future-challenges}

In the previous section we described an SDS-inspired design for smart storage nodes that could enforce privacy rules and make existing storage solutions GDPR-compliant. There are, however, two challenges to be addressed before such a solution can be deployed in practice:
\smallskip

\noindent
\hspace{-2mm}
\colorbox{blue!10}{
\begin{minipage}{0.48\textwidth}

\noindent \textbf{Trusted Execution Environments.}
There are several trust-related challenges in the proposal we made above. First of all, the security of the encrypted data at rest hinges on the assumption that the storage device cannot and will not leak cipher keys. Furthermore, the assumption that all permissions are verified and honored correctly depend on whether the storage node is running the expected software/firmware. For this reason, the Controller has to be able to trust the storage node once it has been powered on and ``booted''. For this, we propose expanding the storage node with a small Trusted Execution Environment that can attest the correctness of the software and hardware contents to the Controller. 

\smallskip

Today, ARM processors can already offer guarantees with the TrustZone extensions but emerging research projects, such as Keystone~\cite{lee2020keystone}, can implement TEEs with custom hardware components. This approach fits the use-case of SDP well, because a small RISC-V or ARM core could be used to load the firmware on the storage node that the controller provides. This firmware, in turn, can be verified not to be able to read out cipher keys to clients, etc. FPGAs have been also proposed to be used as TEEs in project examples such as Cipherbase~\cite{arasu2015transaction} where they perform transaction processing in an always-encrypted database management system. 
\end{minipage}}

\noindent
\hspace{-2mm}
\colorbox{blue!10}{
\begin{minipage}{0.48\textwidth}
\noindent \textbf{Data Tracking Beyond Storage.}
While making sure that the storage layer protects privacy and respects all rules, data misuse can happen at other layers of the application stack as well. There is no guarantee that a buggy or malicious application does not store, for instance, data belonging to one \emph{user} under some other, potentially non-existent, one's data; or that results of processing do not leak personally identifiable information to the outside world. Countering such behavior has been the subject of numerous studies in the context of Information Flow Control, but practical, general purpose, solutions are still not widely available. Even though the SDP approach does not solve this challenge, we believe that at least it makes it easier: On the one hand, removing all decision making and policy interpretation from the storage nodes and moving it into a logically centralized controller allows for better overview of the system. Other monitoring tools might be used to augment the monitoring capability of the controller, achieving this way better coverage. Furthermore, by not allowing storing new tuples into the storage unless they belong to a \emph{user} and \emph{purpose} known to the controller, some misuse scenarios can be limited and be audited after the fact (identifying, for instance, the application that created non-existent user IDs).
\end{minipage}}

\section{Conclusion}

In this paper we painted our vision for a GDPR-compliant storage solution that relies on a control path/data path separation to simplify the complexity of in-storage processing necessary for enforcing privacy rules, and hence making practical implementations possible. We call this approach \emph{Software-Defined Data Protection (SDP)}, inspired by the Software-Defined Storage trend. Thanks to SDP, implementing a processing pipeline in smart storage to ensure complex privacy rules at high bandwidth is within reach. As we sketch in this paper, such functionality could indeed be provided by re-purposing existing building blocks. There are, however, open challenges to be tackled. Importantly, SDP requires storage nodes to be treated as a trusted cloud resource. 

\smallskip

This paper is a call to arms for security and systems researchers to join forces in making privacy protecting, GDPR-compliant, distributed cloud storage a reality.

\bibliographystyle{abbrv}
\bibliography{arxiv}

\begin{thebibliography}{10}

\bibitem{arasu2015transaction}
A.~Arasu, K.~Eguro, M.~Joglekar, R.~Kaushik, D.~Kossmann, and R.~Ramamurthy.
\newblock Transaction processing on confidential data using cipherbase.
\newblock In {\em 2015 IEEE 31st International Conference on Data Engineering},
  pages 435--446. IEEE, 2015.

\bibitem{bailleu2019speicher}
M.~Bailleu, J.~Thalheim, P.~Bhatotia, C.~Fetzer, M.~Honda, and K.~Vaswani.
\newblock $\{$SPEICHER$\}$: Securing lsm-based key-value stores using shielded
  execution.
\newblock In {\em 17th $\{$USENIX$\}$ Conference on File and Storage
  Technologies ($\{$FAST$\}$ 19)}, pages 173--190, 2019.

\bibitem{bauer2018programmable}
W.~Bauer, P.~Holzinger, M.~Reichenbach, S.~Vaas, P.~Hartke, and D.~Fey.
\newblock Programmable hsa accelerators for zynq ultrascale+ mpsoc systems.
\newblock In {\em European Conference on Parallel Processing}, pages 733--744.
  Springer, 2018.

\bibitem{boneh1996revocable}
D.~Boneh and R.~J. Lipton.
\newblock A revocable backup system.
\newblock In {\em USENIX Security Symposium}, pages 91--96, 1996.

\bibitem{caulfield2016cloud}
A.~M. Caulfield, E.~S. Chung, A.~Putnam, H.~Angepat, J.~Fowers, M.~Haselman,
  S.~Heil, M.~Humphrey, P.~Kaur, J.-Y. Kim, et~al.
\newblock A cloud-scale acceleration architecture.
\newblock In {\em 2016 49th Annual IEEE/ACM International Symposium on
  Microarchitecture (MICRO)}, pages 1--13. IEEE, 2016.

\bibitem{ccpa}
CCPA.
\newblock {California Consumer Privacy Act}.
\newblock {\em California Civil Code, Section 1798.100}, Jun 28 2018.

\bibitem{chapmancomputational}
K.~Chapman, M.~Nik, B.~Robatmili, S.~Mirkhani, and M.~Lavasani.
\newblock Computational storage for big data analytics.
\newblock In {\em Proceedings of 10th International Workshop on Accelerating
  Analytics and Data Management Systems (ADMS'19)}, 2019.

\bibitem{do2019programmable}
J.~Do, S.~Sengupta, and S.~Swanson.
\newblock Programmable solid-state storage in future cloud datacenters.
\newblock {\em Communications of the ACM}, 62(6):54--62, 2019.

\bibitem{inproceedings}
B.~Duncan.
\newblock Eu general data protection regulation compliance challenges for cloud
  users.
\newblock 05 2019.

\bibitem{gdpr-regulation}
GDPR.
\newblock Regulation ({EU}) 2016/679 of the {E}uropean {P}arliament and of the
  {C}ouncil of 27 {A}pril 2016 on the protection of natural persons with regard
  to the processing of personal data and on the free movement of such data, and
  repealing {D}irective 95/46.
\newblock {\em Official Journal of the European Union}, 59(1-88), 2016.

\bibitem{hodjat200421}
A.~Hodjat and I.~Verbauwhede.
\newblock A 21.54 gbits/s fully pipelined aes processor on fpga.
\newblock In {\em 12th Annual IEEE Symposium on Field-Programmable Custom
  Computing Machines}, pages 308--309. IEEE, 2004.

\bibitem{istvan2016runtime}
Z.~Istv{\'a}n, D.~Sidler, and G.~Alonso.
\newblock Runtime parameterizable regular expression operators for databases.
\newblock In {\em 2016 IEEE 24th Annual International Symposium on
  Field-Programmable Custom Computing Machines (FCCM)}, pages 204--211. IEEE,
  2016.

\bibitem{istvan2017caribou}
Z.~Istv{\'a}n, D.~Sidler, and G.~Alonso.
\newblock Caribou: intelligent distributed storage.
\newblock {\em Proceedings of the VLDB Endowment}, 10(11):1202--1213, 2017.

\bibitem{istvan2016consensus}
Z.~Istv{\'a}n, D.~Sidler, G.~Alonso, and M.~Vukolic.
\newblock Consensus in a box: Inexpensive coordination in hardware.
\newblock In {\em 13th $\{$USENIX$\}$ Symposium on Networked Systems Design and
  Implementation ($\{$NSDI$\}$ 16)}, pages 425--438, 2016.

\bibitem{jauernig2020trusted}
P.~Jauernig, A.-R. Sadeghi, and E.~Stapf.
\newblock Trusted execution environments: Properties, applications, and
  challenges.
\newblock {\em IEEE Security \& Privacy}, 18(2):56--60, 2020.

\bibitem{jo2016yoursql}
I.~Jo, D.-H. Bae, A.~S. Yoon, J.-U. Kang, S.~Cho, D.~D. Lee, and J.~Jeong.
\newblock Yoursql: a high-performance database system leveraging in-storage
  computing.
\newblock {\em Proceedings of the VLDB Endowment}, 9(12):924--935, 2016.

\bibitem{jun2016bluedbm}
S.-W. Jun, M.~Liu, S.~Lee, J.~Hicks, J.~Ankcorn, M.~King, and S.~Xu.
\newblock Bluedbm: Distributed flash storage for big data analytics.
\newblock {\em ACM Transactions on Computer Systems (TOCS)}, 34(3):1--31, 2016.

\bibitem{kelly2020achieve}
M.~Kelly, E.~Furey, and K.~Curran.
\newblock How to achieve compliance with gdpr article 17 in a hybrid cloud
  environment.
\newblock {\em Sci}, 2(2):22, 2020.

\bibitem{kim2020evanesco}
M.~Kim, J.~Park, G.~Cho, Y.~Kim, L.~Orosa, O.~Mutlu, and J.~Kim.
\newblock Evanesco: Architectural support for efficient data sanitization in
  modern flash-based storage systems.
\newblock In {\em Proceedings of the Twenty-Fifth International Conference on
  Architectural Support for Programming Languages and Operating Systems}, pages
  1311--1326, 2020.

\bibitem{koo2017summarizer}
G.~Koo, K.~K. Matam, I.~Te, H.~K.~G. Narra, J.~Li, H.-W. Tseng, S.~Swanson, and
  M.~Annavaram.
\newblock Summarizer: trading communication with computing near storage.
\newblock In {\em 2017 50th Annual IEEE/ACM International Symposium on
  Microarchitecture (MICRO)}, pages 219--231. IEEE, 2017.

\bibitem{krahn2018pesos}
R.~Krahn, B.~Trach, A.~Vahldiek-Oberwagner, T.~Knauth, P.~Bhatotia, and
  C.~Fetzer.
\newblock Pesos: Policy enhanced secure object store.
\newblock In {\em Proceedings of the Thirteenth EuroSys Conference}, pages
  1--17, 2018.

\bibitem{kraska2019schengendb}
T.~Kraska, M.~Stonebraker, L.~M. Brodie, S.~Servan-Schreiber, and J.~D.
  Weitzner.
\newblock Schengendb - a data protection database proposal.
\newblock {\em Poly/DMAH@VLDB}, pages 24--38, 2019.

\bibitem{kuhring2019specialize}
L.~Kuhring, E.~Garcia, and Z.~Istv{\'a}n.
\newblock Specialize in moderation—building application-aware storage
  services using fpgas in the datacenter.
\newblock In {\em 11th $\{$USENIX$\}$ Workshop on Hot Topics in Storage and
  File Systems (HotStorage 19)}, 2019.

\bibitem{lee2020keystone}
D.~Lee, D.~Kohlbrenner, S.~Shinde, K.~Asanovi{\'c}, and D.~Song.
\newblock Keystone: an open framework for architecting trusted execution
  environments.
\newblock In {\em Proceedings of the Fifteenth European Conference on Computer
  Systems}, pages 1--16, 2020.

\bibitem{li2017kv}
B.~Li, Z.~Ruan, W.~Xiao, Y.~Lu, Y.~Xiong, A.~Putnam, E.~Chen, and L.~Zhang.
\newblock Kv-direct: High-performance in-memory key-value store with
  programmable nic.
\newblock In {\em Proceedings of the 26th Symposium on Operating Systems
  Principles}, pages 137--152, 2017.

\bibitem{8933680}
Z.~S. {Li}, C.~{Werner}, and N.~{Ernst}.
\newblock Continuous requirements: An example using gdpr.
\newblock In {\em 2019 IEEE 27th International Requirements Engineering
  Conference Workshops (REW)}, pages 144--149, 2019.

\bibitem{macedosurvey}
R.~Macedo, J.~Paulo, J.~Pereira, and A.~Bessani.
\newblock A survey and classification of software-defined storage systems.
\newblock {\em ACM Computing Surveys (CSUR)}.

\bibitem{Mazmudar2019MitigatorPP}
M.~Mazmudar.
\newblock Mitigator: Privacy policy compliance using intel sgx.
\newblock 2019.

\bibitem{mitra2006secure}
S.~Mitra and M.~Winslett.
\newblock Secure deletion from inverted indexes on compliance storage.
\newblock In {\em Proceedings of the second ACM workshop on Storage security
  and survivability}, pages 67--72, 2006.

\bibitem{ponnapalli2019scalable}
S.~Ponnapalli, A.~Shah, A.~Tai, S.~Banerjee, V.~Chidambaram, D.~Malkhi, and
  M.~Wei.
\newblock Scalable and efficient data authentication for decentralized systems.
\newblock {\em arXiv preprint arXiv:1909.11590}, 2019.

\bibitem{raju2018mlsm}
P.~Raju, S.~Ponnapalli, E.~Kaminsky, G.~Oved, Z.~Keener, V.~Chidambaram, and
  I.~Abraham.
\newblock mlsm: Making authenticated storage faster in ethereum.
\newblock In {\em 10th $\{$USENIX$\}$ Workshop on Hot Topics in Storage and
  File Systems (HotStorage 18)}, 2018.

\bibitem{Rios2019ServiceLA}
E.~Rios, E.~Iturbe, X.~Larrucea, M.~Rak, W.~Mallouli, J.~Dominiak,
  V.~Munt{\'e}s, P.~Matthews, and L.~Gonzalez.
\newblock Service level agreement-based gdpr compliance and security assurance
  in (multi)cloud-based systems.
\newblock {\em IET Software}, 13:213--222, 2019.

\bibitem{samsungsmartssd}
Samsung.
\newblock Samsung smartssd product brief.
\newblock
  https://www.nimbix.net/wp-content/uploads/2020/02/SmartSSD\_ProductBrief\_12.pdf,
  2020.

\bibitem{schwarzkopf2019position}
M.~Schwarzkopf, E.~Kohler, M.~F. Kaashoek, and R.~Morris.
\newblock Position: Gdpr compliance by construction.
\newblock In {\em Heterogeneous Data Management, Polystores, and Analytics for
  Healthcare}, pages 39--53. Springer, 2019.

\bibitem{sen2014bootstrapping}
S.~Sen, S.~Guha, A.~Datta, S.~K. Rajamani, J.~Tsai, and J.~M. Wing.
\newblock Bootstrapping privacy compliance in big data systems.
\newblock In {\em 2014 IEEE Symposium on Security and Privacy}, pages 327--342.
  IEEE, 2014.

\bibitem{shah2019analyzing}
A.~Shah, V.~Banakar, S.~Shastri, M.~Wasserman, and V.~Chidambaram.
\newblock Analyzing the impact of $\{$GDPR$\}$ on storage systems.
\newblock In {\em 11th $\{$USENIX$\}$ Workshop on Hot Topics in Storage and
  File Systems (HotStorage 19)}, 2019.

\bibitem{10.14778/3384345.3384354}
S.~Shastri, V.~Banakar, M.~Wasserman, A.~Kumar, and V.~Chidambaram.
\newblock Understanding and benchmarking the impact of gdpr on database
  systems.
\newblock {\em Proc. VLDB Endow.}, 13(7):1064–1077, Mar. 2020.

\bibitem{shastri2019seven}
S.~Shastri, M.~Wasserman, and V.~Chidambaram.
\newblock The seven sins of personal-data processing systems under
  $\{$GDPR$\}$.
\newblock In {\em 11th $\{$USENIX$\}$ Workshop on Hot Topics in Cloud Computing
  (HotCloud 19)}, 2019.

\bibitem{thereska2013ioflow}
E.~Thereska, H.~Ballani, G.~O'Shea, T.~Karagiannis, A.~Rowstron, T.~Talpey,
  R.~Black, and T.~Zhu.
\newblock Ioflow: a software-defined storage architecture.
\newblock In {\em Proceedings of the Twenty-Fourth ACM Symposium on Operating
  Systems Principles}, pages 182--196, 2013.

\bibitem{upadhyaya2015automatic}
P.~Upadhyaya, M.~Balazinska, and D.~Suciu.
\newblock Automatic enforcement of data use policies with datalawyer.
\newblock In {\em Proceedings of the 2015 ACM SIGMOD International Conference
  on Management of Data}, pages 213--225, 2015.

\bibitem{wang2019riverbed}
F.~Wang, R.~Ko, and J.~Mickens.
\newblock Riverbed: enforcing user-defined privacy constraints in distributed
  web services.
\newblock In {\em 16th $\{$USENIX$\}$ Symposium on Networked Systems Design and
  Implementation ($\{$NSDI$\}$ 19)}, pages 615--630, 2019.

\bibitem{zhu2005fossilized}
Q.~Zhu and W.~W. Hsu.
\newblock Fossilized index: The linchpin of trustworthy non-alterable
  electronic records.
\newblock In {\em Proceedings of the 2005 ACM SIGMOD international conference
  on Management of data}, pages 395--406, 2005.

\end{thebibliography}

\end{document}